\documentclass{elsart}
\usepackage{epsfig}

\begin{document}

\begin{frontmatter}
\title{Characterization and Modeling of weighted networks}

\author{Marc Barth{\'e}lemy$^1$, Alain Barrat$^2$, Romualdo Pastor-Satorras$^3$,}
\author{and Alessandro Vespignani$^2$}

\address{
$^1$ CEA-Centre d'Etudes de
  Bruy{\`e}res-le-Ch{\^a}tel, D\'epartement de Physique Th\'eorique et
  Appliqu\'ee BP12, 91680 Bruy{\`e}res-Le-Ch{\^a}tel, France\\
$^2$ Laboratoire de Physique Th{\'e}orique (UMR du CNRS
 8627), B\^atiment 210, Universit{\'e} de Paris-Sud 91405 Orsay, France\\
$^3$ Departament de F\'{i}sica i Enginyeria Nuclear, Universitat
Polit\`{e}cnica de Catalunya, Campus Nord, M\`{o}dul B4, 08034
Barcelona, Spain}

\begin{abstract}
  
  We review the main tools which allow for the statistical
  characterization of weighted networks. We then present two case
  studies, the airline connection network and the scientific
  collaboration network, which are representative of critical
  infrastructures and social systems, respectively. The main empirical
  results are (i) the broad distributions of various quantities and
  (ii) the existence of weight-topology correlations. These
  measurements show that weights are relevant and that in general the
  modeling of complex networks must go beyond topology. We review a
  model which provides an explanation for the features observed in
  several real-world networks. This model of weighted network
  formation relies on the dynamical coupling between topology and
  weights, considering the rearrangement of weights when new links are
  introduced in the system.

\vspace{0.25cm}
\end{abstract}

\end{frontmatter}

\vspace{-0.3cm}

\section{Introduction}

\vspace{-0.3cm}

Networked structures arise in a wide array of different contexts such
as technological and transportation infrastructures, social phenomena,
and biological systems.  These highly interconnected systems have
recently been the focus of a great deal of attention that has
uncovered and characterized their topological
complexity~\cite{Barabasi:2002,mdbook,psvbook,Amaral:2000}. Along with
a complex topological structure, real networks display a large
heterogeneity in the capacity and intensity of the connections---the
weight of the links. For example, in ecology the diversity of the
predator-prey interaction is believed to be a critical ingredient of
ecosystems stability~\cite{Pimm,Krause:2003}, and in social systems,
the weight of interactions is very important in the characterization
of the corresponding networks~\cite{Granovetter}. Similarly, the
Internet traffic~\cite{psvbook} or the number of passengers in the
airline network~\cite{Amaral:2000,luisair,Barrat:2004a} are crucial
quantities in the study of these systems.

\vspace{-0.3cm}

In this paper we review a set of metrics combining weighted and
topological observables that allows to characterize the complex
statistical properties of the strength of edges and vertices and to
investigate the correlations among weighted quantities and the
underlying topological structure. Specifically, we present results on
the scientific collaboration network and the world-wide
air-transportation network, which are representative examples of
social and large infrastructure systems, respectively. The measures on
weighted networks~\cite{Barrat:2004a,Li:2003a,Li:2003b,Garla:2003}
have shown that they can exhibit additional complex properties such as
broad distributions and non-trivial correlations of weights that do
not find an explanation just in terms of the underlying topological
structure. The heterogeneity in the intensity of connections may thus
be very important in real-world systems and cannot be overlooked in
their description. Motivated by these observations, we review also a
model for weighted networks we have recently
proposed~\cite{Barrat:2004b}, which naturally produces topology-weight
correlations and broad distributions for various quantities.

\vspace{-0.3cm}
\section{Tools for the characterization of weighted networks}

\vspace{-0.3cm}

We briefly review the different tools which allow for a first statistical
characterization of weighted complex networks.

\begin{itemize}

\item{} {\it Weights}
  
  The properties of a graph can be expressed via its adjacency matrix
  $a_{ij}$, whose elements take the value $1$ if an edge connects the
  vertex $i$ to the vertex $j$, and $0$ otherwise (with $i,j=
  1,...,N$, where $N$ is the size of the network). Weighted networks
  are usually described by a matrix $w_{ij}$ specifying the weight on
  the edge connecting the vertices $i$ and $j$ ($w_{ij}=0$ if the
  nodes $i$ and $j$ are not connected). In the following we will
  consider only the
  case of symmetric positive weights $w_{ij}=w_{ji}\geq 0$.\\

\item{} {\it Connectivity and weight distributions}

The standard topological characterization of networks is obtained by
the analysis of the probability distribution $P(k)$ that a vertex has
degree $k$. Complex networks often exhibits a power-law degree
distribution $P(k)\sim k^{-\gamma}$ with $2\leq\gamma\leq
3$. Similarly, a first characterization of weights is obtained by the
distribution $P(w)$ that any given edge has weight $w$.\\

\item{} {\it Weighted connectivity: Strength}

Along with the degree of a node, a very significative measure of the
network properties in terms of the actual weights is obtained by
looking at the vertex {\it strength} $s_i$ defined
as~\cite{Yook:2001,Barrat:2004a,Onnela:2003}
\begin{equation}
s_i=\sum_{j\in{\mathcal V}(i)}w_{ij},
\end{equation}
where the sum runs over the set ${\mathcal V}(i)$ of neighbors of $i$.
The strength of a node integrates the information both about its
connectivity and the importance of the weights of its links, and can
be considered as the natural generalization of the connectivity. When
the weights are independent from the topology, we obtain that the
strength of the vertices of degree $k$ is $s(k)\simeq \langle w \rangle k$ where $\langle w\rangle $
is the average weight. In the presence of correlations we obtain in
general $s(k)\simeq Ak^\beta$ with $\beta=1$ and $A\neq \langle w\rangle $ or
$\beta>1$.\\

\item{} {\it Weighted clustering}

The clustering coefficient $c_i$
  measures the local cohesiveness and is defined for any vertex $i$ as
  the fraction of connected neighbors of $i$ \cite{Watts:98}. The
  average clustering coefficient $C=N^{-1}\sum_i c_i$ thus expresses the
  statistical level of cohesiveness measuring the global density of
  interconnected vertex triplets in the network.  Further information
  can be gathered by inspecting the average clustering coefficient
  $C(k)$ restricted to the class of vertices with degree $k$.  The
  topological clustering, however, does not take into account the fact
  that some neighbors are more important than others. In order to
  solve this incongruity we introduce a measure of the clustering that
  combines the topological information with the weight distribution of
  the network.  The {\em weighted clustering coefficient} is defined
  as~\cite{Barrat:2004a}
\begin{equation}
  c^w(i)=\frac{1}{s_i(k_i-1)} \sum_{j, h}
  \frac{(w_{ij}+w_{ih})}{2} a_{ij}a_{ih}a_{jh}.
\end{equation}
This quantity $c^w(i)$ counts for each triple formed in the
neighborhood of the vertex $i$ the weight of the two participating
edges of the vertex $i$. In this way we are not just considering the
number of closed triangles in the neighborhood of a vertex but also
their total relative weight with respect to the vertex' strength. The
factor $s_i(k_i-1)$ is a normalization factor and ensures that $0\leq
c_{i}^w\leq 1$. Consistently, the $c_{i}^w$ definition recovers the
topological clustering coefficient in the case that $w_{ij}=const$. It
is customary to define $C^w$ and $C^w(k)$ as the weighted clustering
coefficient averaged over all vertices of the network and over all
vertices with degree $k$, respectively. In the case of a large
randomized network (lack of correlations) it is easy to see that
$C^w=C$ and $C^w(k)=C(k)$. In real weighted networks, however, we can
face two opposite cases. If $C^w > C$, we are in presence of a network
in which the interconnected triples are more likely formed by the
edges with larger weights. On the contrary, $C^w<C$ signals a network
in which the topological clustering is generated by edges with low
weight. In this case it is obvious that the clustering has a minor
effect in the organization of the network since the largest part of
the interactions (traffic, frequency of the relations, etc.) is
occurring on edges not belonging to interconnected triples. The same
may happen for $C^w(k)$, for which it is also possible to analyze the
variations with respect to the degree class $k$.\\

\item{} {\it Weighted assortativity: Affinity}

 Another quantity
used to probe the networks' architecture is
the average degree of nearest neighbors of a vertex $i$ 
\begin{equation}
k_{nn,i}=\frac{1}{k_i}\sum_{j \in \mathcal{V}(i)}  k_j \ ,
\end{equation}
where the sum runs on the nearest neighbors vertices of each vertex
$i$.  From this quantity a convenient measure 
to investigate the behavior of the degree
correlation function is obtained by the  average degree of the 
nearest neighbors, $k_{nn}(k)$, for vertices of degree
$k$\cite{alexei}. 
In the presence of correlations, the behavior of
$k_{nn}(k)$ identifies two general classes of networks. If $k_{nn}(k)$
is an increasing function of $k$, vertices with high degree have a
larger probability to be connected with large degree vertices. This
property is referred in physics and social sciences as {\em
  assortative mixing} \cite{assortative}. On the contrary, a
decreasing behavior of $k_{nn}(k)$ defines {\em disassortative
  mixing}, in the sense that high degree vertices have a majority of
neighbors with low degree, while the opposite holds for low degree
vertices.

In the case of weighted networks an appropriate characterization
of the assortative behavior is obtained by  the {\em
weighted average nearest neighbors degree}, defined as
\begin{equation}
  k^w_{nn,i}=\frac{1}{s_i}\sum_{j=1}^N a_{ij} w_{ij} k_j.
\end{equation}
In this case, we perform a local weighted average of the nearest
neighbor degree according to the normalized weight of the connecting
edges, $w_{ij} / s_i$.  This definition implies that $k^w_{nn,i}>
k_{nn,i}$ if the edges with the larger weights are pointing to the
neighbors with larger degree and $k^w_{nn,i}< k_{nn,i}$ in the
opposite case. The $k^w_{nn,i}$ function thus measures the effective
{\em affinity} to connect with high or low degree neighbors according
to the magnitude of the actual interactions.  As well, the behavior of
the function $k^w_{nn}(k)$ (defined as the average of $k^w_{nn,i}$
over all vertices with degree $k$), marks the weighted assortative or
disassortative properties considering the actual interactions among
the system's elements.\\

\item{} {\it Disparity}

For a given node $i$ with connectivity $k_i$ and strength $s_i$
different situations can arise. All weights $w_{ij}$ can be of the
same order $s_i/k_i$. In contrast, the most heterogeneous situation is
obtained when one weight dominates over all the others. A simple way
to measure this ``disparity'' is given by the quantity $Y_2$
introduced in other context~\cite{Derrida:1987,Barthelemy:2003}
\begin{equation}
Y_2(i)=\sum_{j\in{\mathcal V}(i)}\left[\frac{w_{ij}}{s_i}\right]^2
\end{equation}
If all weights are of the same order then $Y_2\sim 1/k_i$ (for $k_i\gg
1$) and if a small number of weights dominate then $Y_2$ is of the
order $1/n$ with $n$ of order unity. This quantity was recently used
for metabolic networks~\cite{Barabasi:2004} which showed that for
these networks one can identify dominant reactions.

\end{itemize}

\vspace{-0.3cm}

\section{Empirical results}

\vspace{-0.3cm}

\subsection{Weighted networks data}

\vspace{-0.3cm}

Prototypical examples of weighted networks can be found in the
world-wide airport network (WAN)~\cite{luisair,Barrat:2004a} and the
scientific collaboration network (SCN)~\cite{newman01,vicsek}.  In the
airport network each given weight $w_{ij}$ is the number of available
seats on direct flight connections between the airports $i$ and
$j$. For the WAN, we analyze the International Air Transportation
Association (IATA)\footnote{http://www.iata.org.} database containing
the world list of airports pairs connected by direct flights and the
number of available seats on any given connection for the year
2002. The resulting air-transportation graph comprises $N=3880$
vertices denoting airports and $E=18810$ edges accounting for the
presence of a direct flight connection.  The average degree of the
network is $\langle k \rangle =2E/N=9.70$, while the maximal degree is
$318$.

\vspace{-0.3cm}

In the SCN the nodes are identified with authors and the weight
depends on the number of co-authored
papers~\cite{newman01,Barrat:2004a}. We consider the network of
scientists who have authored manuscripts submitted to the e-print
archive relative to condensed matter physics
(http://xxx.lanl.gov/archive/cond-mat) between 1995 and
1998. Scientists are identified with nodes and an edge exists between
two scientists if they have co-authored at least one paper. The
resulting connected network has $N=12722$ nodes, with an average
degree (i.e. average number of collaborators) $\langle k \rangle=6.28$
and maximal degree $97$. For the SCN we follow the definition of
weight introduced in Ref.~\cite{newman01}: The intensity $w_{ij}$ of
the interaction between two collaborators $i$ and $j$ is defined as
$w_{ij} = \sum_p \ \delta_i^p \delta_j^p/(n_p -1)$ where the index $p$
runs over all papers, $n_p$ is the number of authors of the paper $p$,
and $\delta_i^p$ is $1$ if author $i$ has contributed to paper $p$,
and $0$ otherwise. This definition seems to be rather objective and
representative of the scientific interaction: It is large for
collaborators having many papers in common but the contribution to the
weight introduced by any given paper is inversely proportional to the
number of authors.

\vspace{-0.6cm}
\subsection{Empirical results}

\vspace{-0.3cm}

\subsubsection{Topological properties}

The topological properties of the SCN network and other similar
networks of scientific collaborations have been studied in
Ref.~\cite{newman01} and we report on Fig.~(\ref{fig1}A) the
connectivity distribution showing a relatively broad law.

\vspace{-0.3cm}

As shown in Fig.~(\ref{fig1}B), the topology of the WAN exhibits both
small-world and scale-free properties as already observed in different
dataset analyses \cite{Li:2003a,luisair}. In particular, the average
shortest path length, measured as the average number of edges
separating any two nodes in the network, shows the value
$\langle\ell\rangle=4.37$, very small compared to the network size
$N$. The degree distribution, on the other hand, takes the form $P(k)=
k^{-\gamma}f(k/k_x)$, where $\gamma\simeq2.0$ and $f(k/k_x)$ is an
exponential cut-off function that finds its origin in physical
constraints on the maximum number of connections that a single airport
can handle \cite{Amaral:2000,luisair}. The airport connection graph is
therefore a clear example of heterogeneous network showing scale-free
properties on a definite range of degree values.

\vspace{-0.5cm}

\subsubsection{Strength distribution}

\begin{figure}
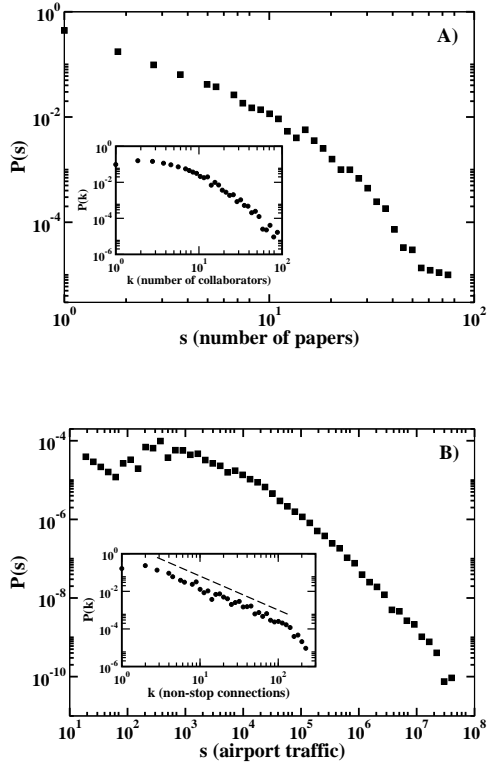

\epsfxsize=2.5in
\centerline{\epsffile{fig1a.eps}}
\vspace*{1.0cm}
\centerline{\epsffile{fig1b.eps}}
\caption{ {\bf A} Degree and strength distribution 
  in the scientific collaboration network. The degree $k$ corresponds
  to the number of co-authors of each scientist and the strength
  represent its total number of publications. The distributions are
  heavy-tailed even if it is not possible to distinguish a definite
  functional form.
  {\bf B} The same distributions for the world-wide airport network.
  The degree is the number of non-stop connections to other airports
  and the strength is the total number of passengers handled by any
  given airport. In this case, the degree distribution can be
  approximated by the power-law behavior $P(k)\sim k ^{-\gamma}$ with
  $\gamma=1.8\pm 0.2$. The strength distribution has a heavy-tail extending
  over more than four orders of magnitude.}
\label{fig1}
\end{figure}

The probability distribution $P(s)$ that a vertex has strength $s$ is
heavy tailed in both networks and the functional behavior exhibits
similarities with the degree distribution $P(k)$ (see
Fig.~\ref{fig1}). A precise functional description of the heavy-tailed
distributions may be very important in understanding the network
evolution and will be deferred to future analysis. This behavior is
not unexpected since it is plausible that the strength $s_i$ increases
with the vertex degree $k_i$, and thus the slow decaying tail of
$P(s)$ stems directly from the very slow decay of the degree
distribution.

\subsubsection{Topology-weight correlations}

\begin{figure}
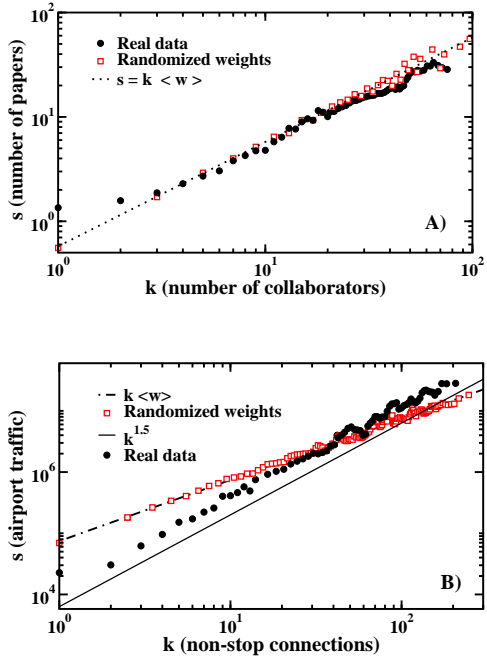

\epsfxsize=2.5in
\centerline{\epsffile{fig2a.eps}}
\vspace*{0.8cm}
\centerline{\epsffile{fig2b.eps}}
\vspace*{0.3cm}
\caption{  Average strength $s(k)$ as function of the degree $k$ of nodes. 
  {\bf A} In the scientific collaboration network the real data are
  very similar to those obtained in a randomized weighted network.
  Only at very large $k$ values it is possible to observe a slight
  departure from the expected linear behavior.  {\bf B} In the world
  airport network real data follow a power-law behavior with exponent
  $\beta=1.5\pm0.1$. This denotes anomalous correlations between the
  traffic handled by an airport and the number of its connections.}
\label{fig2}
\end{figure}

In Fig.~\ref{fig2} we report the behavior obtained for both the real
weighted networks and their randomized versions, generated by a random
re-distribution of the actual weights on the existing topology of the
network.  For the SCN the curves are very similar and well fitted by
the uncorrelated approximation $s(k)= \langle w \rangle k$.  Strikingly, this is
not the case of the WAN.  Fig.~\ref{fig2}B clearly shows a very
different behavior for the real data set and its randomized version.
In particular, the power-law fit for the real data gives an
``anomalous'' exponent $\beta_{\rm WAN}=1.5\pm 0.1$.  This implies that the
strength of vertices grows faster than their degree, i.e.  the weight
of edges belonging to highly connected vertices tends to have a value
higher than the one corresponding to a random assignment of weights.
This denotes a strong correlation between the weight and the
topological properties in the WAN, where the larger is an airport, the
more traffic it can handle.

The fingerprint of these correlations is also observed in the
dependence of the weight $w_{i j}$ on the degrees of the end point
nodes $k_i$ and $k_j$. For the WAN the behavior of the average weight
as a function of the end points degrees can be well approximated by a
power-law dependence $\langle w_{ij} \rangle \sim(k_ik_j)^{\theta}$
with an exponent $\theta=0.5\pm 0.1$.  This exponent can be related to
the $\beta$ exponent by noticing that $s(k) \sim k(kk_j)^{\theta}$,
resulting in $\beta=1+\theta$, if the topological correlations between
the degree of connected vertices can be neglected. This is indeed the
case of the WAN where the above scaling relation is well satisfied by
the numerical values provided by the independent measurements of the
exponents. In the SCN, instead, $\langle w_{ij} \rangle$ is almost
constant for over two decades confirming a general lack of
correlations between the weights and the vertices degree. In this case
$\theta=0$ and the relation $\beta=1+\theta$ also holds.


\subsubsection{Weighted clustering and assortativity}

We present the results~\cite{Barrat:2004a} obtained for both the SCN
(see Fig.\ref{figscn}) and the WAN (see Fig.\ref{figwan})
by comparing the regular topological quantities with the
weighted ones introduced above. In the figures we report the relative 
difference of the values obtained for the topological and weighted
quantities. It is striking to observe that for large degree values 
we observe up to 100\% relative difference signaling a strong 
difference in the clustering and correlation properties if we take
into account the weighted nature of the networks.

\begin{itemize}

\item{} {\it SCN}

\begin{itemize}
\item (i) The measurements indicate that the SCN has a
monotonously decaying spectrum $C(k)$. This implies that hubs present
a much lower clustered neighborhood than low degree vertices which can be
interpreted as the evidence that authors with few
collaborators usually work within a well defined research group in
which all the scientists collaborate together (high clustering).
Authors with a large degree, however, collaborate with different
groups and communities which on their turn do not have often
collaborations, thus creating a lower clustering coefficient.

\item (ii) The inspection of $C^w(k)$ shows generally that
for $k\geq 10$ the weighted clustering coefficient is larger than the
topological one. This implies that authors with many collaborators
tend to publish more papers with interconnected groups of co-authors
and is a signature of the fact that influential scientists form stable
research groups where the largest part of their production is
obtained.

\item (iii) Furthermore, the SCN exhibits an assortative behavior in
agreement with the general evidence that social networks are usually
denoted by a strong assortative character \cite{assortative}. Finally,
the assortative properties find a clearcut confirmation in the
weighted analysis with a $k^w_{nn}(k)$ strikingly growing as a
power-law as a function of $k$.
\end{itemize}
\begin{figure}
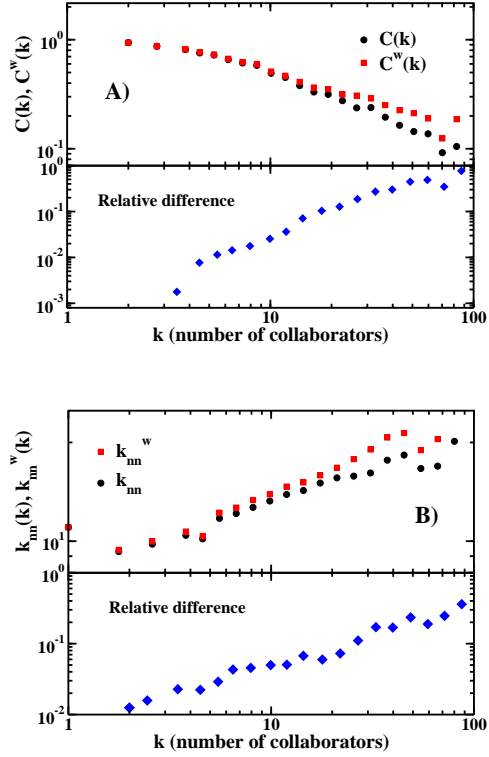

\epsfxsize=2.5in
\centerline{\epsffile{figclustSCN.eps}}
\vspace*{0.8cm}
\centerline{\epsffile{figknnSCN.eps}}
\vspace*{0.3cm}
\caption{ Comparison of topological and weighted quantities for the SCN.
  {\bf A} The weighted clustering separates from  the topological
  one around $k\geq 10$. This marks a difference for authors with 
  larger number of collaborators. 
  {\bf B} The assortative behavior is enhanced in the weighted
  definition of the average nearest neighbors degree. It is worth
remarking the large relative difference (up to 50-100\%) of the weighted and
topological quantities (lower part of the figures).}
\label{figscn}
\end{figure}

\item{} {\it WAN}

A different picture is found in the WAN, where the weighted analysis
provides a richer and somehow different scenario.

\begin{itemize}
\item (i) This network also
shows a decaying $C(k)$, consequence of the role of large airports
that provide non-stop connections to very far destinations on an
international and intercontinental scale. These destinations are
usually not interconnected among them, giving rise to a low clustering
coefficient for the hubs. 
\begin{figure}
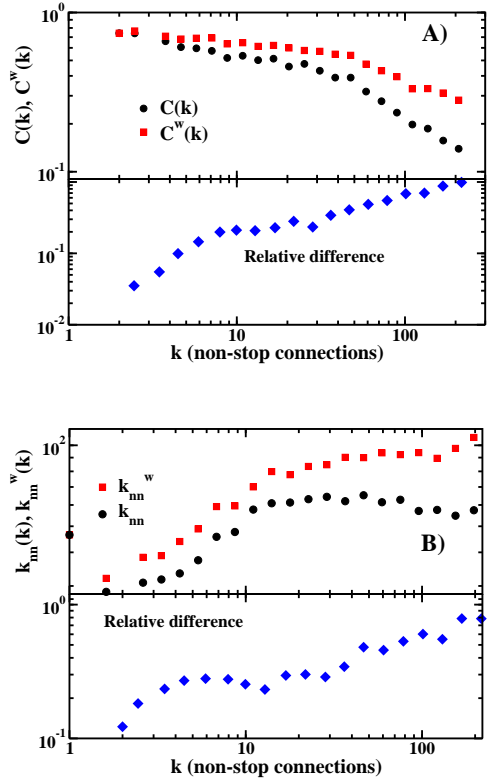

\epsfxsize=2.5in
\centerline{\epsffile{figclustairports.eps}}
\vspace*{0.8cm}
\centerline{\epsffile{figknnairports.eps}}
\vspace*{0.3cm}
\caption{ Topological and weighted quantities for the WAN.
  {\bf A} The weighted clustering coefficient is larger than 
  the topological one in the whole degree spectrum.
  {\bf B} $k_{nn}(k)$ is reaching a plateau for  $k>10$ 
  denoting the absence of marked topological correlations.
  On the contrary $k^w_{nn}(k)$ exhibits  a more definite 
  assortative behavior. Also in this case the relative difference 
  of the weighted and topological quantities are of the order of 
 100\% (lower part of the figures).}
\label{figwan}
\end{figure}
\item (ii) We find, however, that $C^w/C\simeq 1.1$,
indicating an accumulation of traffic on interconnected groups of
vertices. 

\item (iii) The weighted clustering coefficient $C^w(k)$ has much more
limited variation in the whole spectrum of $k$. This implies that high
degree airports have a progressive tendency to form interconnected
groups with high traffic links, thus balancing the reduced topological
clustering. Since high traffic is associated to hubs, we have a
network in which high degree nodes tend to form cliques with nodes
with equal or higher degree, the so-called {\em rich-club phenomenon}
\cite{richclub}.  

\item (iv) The topological $k_{nn}(k)$ does show
an assortative behavior only at small degrees.  For $k>10$,
$k_{nn}(k)$ approaches a constant value, a fact revealing an
uncorrelated structure in which vertices with very different degrees
have a very similar neighborhood. The analysis of the weighted
$k^w_{nn}(k)$, however, exhibits a pronounced assortative behavior in
the whole $k$ spectrum, providing a different picture in which high
degree airports have a larger affinity for other large airports where
the major part of the traffic is directed.
\end{itemize}

\end{itemize}

\section{Modeling weighted networks}



Previous approaches to the modeling of weighted networks focused on
growing topologies where weights were assigned statically, i.e. once
and for ever, with different rules related to the underlying
topology~\cite{Yook:2001,zheng03} (see also
\cite{Krapivsky:2004} for a more recent model). 
These mechanisms, however, overlook
the dynamical evolution of weights according to the topological
variations. We can illustrate this point in the case of the airline
network. If a new airline connection is created between two airports
it will generally provoke a modification of the existing traffic of
both airports. In general, it will increase the traffic activity
depending on the specific nature of the network and on the local
dynamics.  In the following, we review a model that takes into account
the coupled evolution in time of topology and weights.  Instead of
drawing randomly the weights, an alternative consists in coupling the
evolution of the weights and of the topology and allowing the
dynamical evolution of weights during the growth of the system.  This
mimics the evolution and reinforcements of interactions in natural and
infrastructure networks.


The model dynamics starts from an initial seed of $N_0$ vertices
connected by links with assigned weight $w_0$.  At each time step, a
new vertex $n$ is added with $m$ edges (with initial weight $w_0$)
that are randomly attached to a previously existing vertex $i$
according to the probability distribution
\begin{equation}
\Pi_{n\to i}=\frac{s_i}{\sum_j s_j}.
\label{sdrive}
\end{equation}
This rule of ``busy get busier'' relaxes the usual degree preferential
attachment, focusing on a strength driven attachment in which new
vertices connect more likely to vertices handling larger weights and
which are more central in terms of the strength of interactions. This
weight driven attachment (Eq.~(\ref{sdrive})) appears to be a
plausible mechanism in many networks (see also
\cite{Dorogovtsev:2004} for an alternative mechanism
in which {\em edges} with large weight are chosen for
preferential attachment). In the Internet new routers
connect to more central routers in terms of bandwidth and traffic
handling capabilities and in the airport networks new connections are
generally established to airports with a large passenger traffic. Even
in the SCN this mechanism might play a role since an author with more
co-authored papers is more visible and open to further
collaborations. 

\vspace{-0.3cm}

\begin{figure}
\epsfxsize=2.5in
\centerline{\epsffile{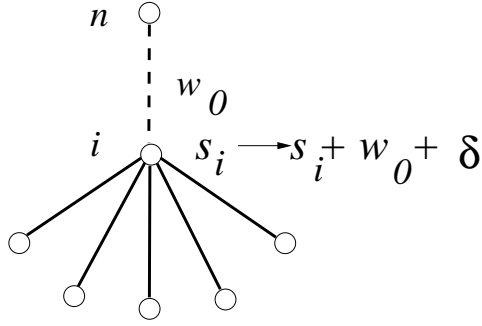}}
\caption{ Illustration of the construction rule. A new node $n$
connects to a node $i$ with probability proportional to 
$ s_i/\sum_j s_j$. The weight of the new edge is $w_0$ and the 
total weight on the existing edges connected to $i$ is modified by an amount equal
to $\delta$. }
\label{fig:rule}
\end{figure}

The presence of the new edge $(n,i)$ will introduce
variations of the existing weights across the network. In particular,
we consider the local rearrangements of weights between $i$ and its
neighbors $j\in{\mathcal V}(i)$ according to the simple rule
\begin{equation}
w_{ij}\to w_{ij}+\Delta w_{ij},
\end{equation}
where 
\begin{equation}
\Delta w_{ij}=\delta\frac{w_{ij}}{s_i}.
\end{equation} 
This rule considers that the establishment of a new edge of weight
$w_0$ with the vertex $i$ induces a total increase of traffic $\delta$
that is proportionally distributed among the edges departing from the
vertex according to their weights (see Fig.~\ref{fig:rule}), yielding
$s_i\to s_i+\delta+w_0$.  At this stage, it is worth remarking that
while we will focus on the simplest model with $\delta=const$,
different choices of $\Delta w_{ij}$ with heterogeneous $\delta_i$ or
depending on the specific properties of each vertex ($w_{ij}, k_i,
s_i$) can be considered~\cite{Barrat:2004c,Barrat:2004d,Pandya:2004}.
Finally, after the weights have been updated the growth process is
iterated by introducing a new vertex with the corresponding
re-arrangement of weights.

\vspace{-0.3cm}

The model depends only on the dimensionless parameter $\delta$
(rescaled by $w_0$), that is the fraction of weight which is `induced'
by the new edge onto the others. According to the value of $\delta$,
different scenarios are possible. If the induced weight is
$\delta\approx 1$ we mimic situations in which an appreciable
fraction of traffic generated by the new connection will be dispatched
in the already existing connections. This is plausible in the airport
networks where the transit traffic is rather relevant in hubs.  In the
case of $\delta < 1$ we face situations such as the SCN where it is
reasonable to consider that the birth of a new collaboration
(co-authorship) is not triggering a more intense activity on previous
collaborations.  Finally, $\delta>1$ is an extreme case in which a new
edge generates a sort of multiplicative effect that is bursting the
weight or traffic on neighbors.

\vspace{-0.3cm}

The network's evolution can be inspected analytically by studying the
time evolution of the average value of $s_i(t)$ and $k_i(t)$ of the
$i$-th vertex at time $t$, and by relying on the continuous
approximation that treats $k$, $s$ and the time $t$ as continuous
variables~\cite{Barabasi:2002,mdbook,Barrat:2004b}. The behavior of
the strength and the connectivity are easily obtained and one has
in the long time limit 
\begin{equation}
s_i(t)\simeq(2\delta+1)k_i(t)
\label{eq_k.vs.s}
\end{equation}
This proportionality relation $s\sim k$ implies $\beta=1$ but the
prefactor is different from $\langle w\rangle $ which indicates the existence of
correlations between topology and weights. This relation
(\ref{eq_k.vs.s}) is particularly relevant since it states that the
weight-driven dynamics generates in Eq.~(\ref{sdrive}) an effective
degree preferential attachment that is parameter independent.  This
highlights an alternative microscopic mechanism accounting for the
presence of the preferential attachment dynamics in growing
networks. 

\vspace{-0.3cm}

The behavior of the various statistical distribution can be easily
computed and one obtains in the large time limit $P(k)\sim
k^{-\gamma}$ and $P(s)\sim s^{-\gamma}$ with
\begin{equation}
\gamma=\frac{4\delta+3}{2\delta+1}.
\end{equation}
This result shows that the obtained graph is a
scale-free network described by an exponent $\gamma\in[2,3]$ that
depends on the value of the parameter $\delta$.  In particular, when
the addition of a new edge doesn't affect the existing weights
($\delta=0$), the model is topologically equivalent to the
Barab\'asi-Albert model \cite{Barabasi:1999} and the value $\gamma=3$ is
recovered.  For larger values of $\delta$ the distribution is
progressively broader with $\gamma\to 2 $ when $\delta\to \infty$.
This indicates that the weight-driven growth generates scale-free
networks with exponents varying in the range of values usually
observed in the empirical analysis of networked
structures~\cite{Barabasi:2002,mdbook,psvbook}. Noticeably the
exponents are non-universal and depend only on the parameter $\delta$
governing the microscopic dynamics of weights. The model therefore
proposes a general mechanism for the occurrence of varying power-law
behaviors without resorting on more complicate topological rules and
variations of the basic preferential attachment mechanism.

\vspace{-0.3cm}

Similarly to the previous quantities, it is possible to obtain
analytical expressions for the evolution of weights and the relative
statistical distribution~\cite{Barrat:2004b}. The probability
distribution $P(w)$ is in this case also a power-law $P(w)\sim
w^{-\alpha}$ where $\alpha =2+\frac{1}{\delta}$. The exponent $\alpha$
has large variations as a function of the parameter $\delta$ and
$P(w)$ moves from a delta function for $\delta=0$ to a very slow
decaying power-law with $\alpha=2$ if $\delta\to\infty$. This feature
clearly shows that the weight distribution is extremely sensitive to
changes in the microscopic dynamics ruling the network's growth.

\vspace{-0.3cm}

In summary, the networks generated by the model display power-law
behavior for the weight, degree and strength distributions with
non-trivial exponents depending on the unique parameter defining the
model's dynamics.  These results suggest that the inclusion of weights
in networks modeling naturally explains the diversity of scale-free
behavior empirically observed in real networked
structures. Strikingly, the weight-driven growth recovers an effective
preferential attachment for the topological properties, providing a
microscopic explanation for the ubiquitous presence of this mechanism.

\vspace{-0.3cm}

\section{Conclusions and perspectives}

\vspace{-0.3cm}

A more complete view of complex networks is thus provided by the study
of the interactions defining the links of these systems. The weights
characterizing the various connections exhibit complex statistical
features with highly varying distributions and power-law behavior.  In
particular we have considered the specific examples of the scientific
collaboration and world-wide airport networks where it is possible to
appreciate the importance of the correlations between weights and
topology in the characterization of real networks properties. Indeed,
the analysis of the weighted quantities and the study of the
correlations between weights and topology provide a complementary
perspective on the structural organization of the network that might
be undetected by quantities based only on topological information. The
weighted quantities thus offer a quantitative and general approach to
understanding the complex architecture of real weighted networks. The
empirical results shows that purely topological models are inadequate
and that there is a need for  models going beyond pure
topology. The model we have presented is possibly the simplest one in
the class of weight-driven growing networks.  A novel feature in the
model is the weight dynamical evolution occurring when new vertices
and edges are introduced in the system. This simple mechanism produces
a wide variety of complex and scale-free behavior depending on the
physical parameter $\delta$ that controls the local microscopic
dynamics. While a constant parameter $\delta$ is enough to produce a
wealth of interesting network properties, a natural generalization of
the model consists in considering $\delta$ as a function of the
vertices degree or strength. Similarly, more complicated variations of
the microscopic rules may be implemented to mimic in a detailed
fashion particular networked systems
~\cite{Barrat:2004c,Barrat:2004d,Pandya:2004,hu}. 
In particular, an important feature to be included in the 
description of the airline
network is the geographical embedding of the
network~\cite{Barthelemy:2004}. In this perspective the
present model appears as a general starting point for the realistic
modeling of complex weighted networks.

Acknowledgments. We thank IATA for making the airline commercial
flight database available to us. We also thank M.E.J. Newman for
conceding us the possibility of using the scientific collaboration
network data (see \cite{Newman:site}).  A.B., R.P.-S. and A.V. are
partially funded by the European Commission - Fet Open Project COSIN
IST-2001-33555. R.P.-S.  acknowledges financial support from the
Ministerio de Ciencia y Tecnolog{\'\i}a (Spain) and from the
Departament d'Universitats, Recerca i Societat de la Informaci{\'o},
Generalitat de Catalunya (Spain).


\end{document}